\documentclass[11pt]{article}
\usepackage{moriond,epsfig}

\def\Journal#1#2#3#4{{#1} {\bf #2}, #3 (#4)}

\def\NIMA{{\em Nucl. Instrum. Methods} A}

\def\PLB{{\em Phys. Lett.}  B}
\def\PRL{\em Phys. Rev. Lett.}
\def\PRD{{\em Phys. Rev.} D}

\def\be{\begin{equation}}
\def\ee{\end{equation}}
\def\bea{\begin{eqnarray}}
\def\eea{\end{eqnarray}}

\begin{document}
\vspace*{4cm}
\title{THE SEARCH FOR SOLAR AXIONS IN THE CAST EXPERIMENT}

\author{DONGHWA KANG$^{10, }$ \footnote{attending speaker, e-mail: donghwa.kang@cern.ch},
S. ANDRIAMONJE$^{2}$, V. ARSOV$^{13}$, S. AUNE$^{2}$, D. AUTIERO$^{1}$,\\
F. T. AVIGNONE$^{3}$, K. BARTH$^{1}$, A. BELOV$^{11}$, 
B. BELTR\'AN$^{6}$, H. BR\"AUNINGER$^{5}$,\\
J. M. CARMONA$^{6}$, S. CEBRI\'AN$^{6}$, E. CHESI$^{1}$, 
J. I. COLLAR$^{7}$, R. CRESWICK$^{3}$, T. DAFNI$^{4}$,\\
M. DAVENPORT$^{1}$, L. Di LELLA$^{1}$, C. ELEFTHERIADIS$^{8}$, 
J. ENGLHAUSER$^{5}$, G. FANOURAKIS$^{9}$,\\ 
H. FARACH$^{3}$, E. FERRER$^{2}$, H. FISCHER$^{10}$, J. FRANZ$^{10}$, 
P. FRIEDRICH$^{5}$, T. GERALIS$^{9}$,\\ 
I. GIOMATARIS$^{2}$, S. GNINENKO$^{11}$, N. GOLOUBEV$^{11}$, 
M. D. HASINOFF$^{12}$, F. H. HEINSIUS$^{10}$,\\ 
D. H. H. HOFFMANN$^{4}$, I. G. IRASTORZA$^{2}$, J. JACOBY$^{13}$, 
K. K\"ONIGSMANN$^{10}$, R. KOTTHAUS$^{14}$,\\ 
M. KR$\check{\rm C}$MAR$^{15}$, K. KOUSOURIS$^{9}$, M. KUSTER$^{5}$, 
B. LAKI\'C$^{15}$, C. LASSEUR$^{1}$, A. LIOLIOS$^{8}$,\\
A. LJUBI$\check{\rm C}$I\'C$^{15}$, G. LUTZ$^{14}$, G. LUZ\'ON$^{6}$, 
D. W. MILLER$^{7}$, A. MORALES$^{6}$, J. MORALES$^{6}$,\\
M. MUTTERER$^{4}$, A. NIKOLAIDIS$^{8}$, A. ORTIZ$^{6}$, 
T. PAPAEVANGELOU$^{1}$, A. PLACCI$^{1}$, \\
G. RAFFELT$^{14}$, J. RUZ$^{6}$, H. RIEGE$^{4}$, M. L. SARSA$^{6}$, 
I. SAVVIDIS$^{8}$, P. SERPICO$^{14}$,\\
Y. SEMERTZIDIS$^{4}$, L. STEWART$^{1}$, J. D. VIEIRA$^{7}$, 
J. VILLAR$^{6}$, J. VOGEL$^{10}$, L. WALCKIERS$^{1}$,\\
K. ZACHARIADOU$^{9}$ and K. ZIOUTAS$^{16}$\\
(CAST Collaboration)}

\address{$^{1}$European Organization for Nuclear Research (CERN), Gen\`eve, Switzerland \\ 
$^{2}$DAPNIA, Centre d'\'Etudes Nucl\'eaires de Saclay (CEA-Saclay), Gif-sur-Yvette, France \\ 
$^{3}$Department of Physics and Astronomy, University of South Carolina, Columbia, South Carolina, USA \\ 
$^{4}$GSI-Darmstadt and Institut f\"ur Kernphysik, Technische Universit\"at Darmstadt, Darmstadt, Germany\\ 
$^{5}$Max-Planck-Institut f\"ur Extraterrestrische Physik, Garching, Germany \\
$^{6}$Instituto de F{\'\i}sica Nuclear y Altas Energ{\'\i}as, Universidad de Zaragoza, Zaragoza, Spain \\ 
$^{7}$Enrico Fermi Institute and KICP, University of Chicago, Chicago, Illinois, USA \\
$^{8}$Aristotle University of Thessaloniki, Thessaloniki, Greece \\
$^{9}$National Center for Scientific Research ``Demokritos'', Athens, Greece \\
$^{10}$Albert-Ludwigs-Universit\"at Freiburg, Freiburg, Germany \\
$^{11}$Institute for Nuclear Research (INR), Russian Academy of Sciences, Moscow, Russia \\
$^{12}$Department of Physics and Astronomy, University of British Columbia, Vancouver, Canada \\
$^{13}$Johann Wolfgang Goethe-Universit\"at, Institut f\"ur Angewandte Physik, Frankfurt am Main, Germany \\
$^{14}$Max-Planck-Institut f\"ur Physik (Werner-Heisenberg-Institut), Munich, Germany \\
$^{15}$Rudjer Bo$\check{\rm s}$kovi\'c Institute, Zagreb, Croatia \\
$^{16}$Department of Physics, University of Patras, Patras, Greece}

\maketitle
\abstracts{The CAST (CERN Axion Solar Telescope) experiment at CERN searches for solar 
axions with energies in the keV range. It is possible that axions are produced in the 
core of the sun by the interaction of thermal photons with virtual photons of strong electromagnetic fields.
In this experiment, the solar axions can be reconverted to photons in the transversal field of a 9 Tesla 
superconducting magnet. At both ends of the 10m-long dipole magnet three different X-ray
detectors were installed, which are sensitive in the interesting photon energy range.
Preliminary results from the analysis of the 2004 data are presented:
g$_{a\gamma}<0.9\times10^{-10}$ GeV$^{-1}$ at 95\% C.L. 
for axion masses m$_{a} <$ 0.02 eV.
At the end of 2005, data started to be taken with a buffer gas in the magnet pipes 
in order to extend the sensitivity to axion masses up to 0.8 eV.}

\section{Introduction}
In the standard model the QCD Lagrangian \cite{Quigg} can be written
with a gluon interaction term violating 
charge conjugation times parity (CP) and time reversal (T)
\begin{equation}
L_{QCD} = L_{pert} + \overline{\theta} {g^{2} \over 32\pi^{2}}
G_{a}^{\mu\nu} \widetilde{G}_{\mu\nu}^{a} ,
\label{Lqcd}
\end{equation}
where $G_{a}^{\mu\nu}$ is the gluon field strength tensor, 
$\widetilde{G}^{a}_{\mu\nu}$ is the corresponding dual tensor,
and the parameter $\overline{\theta}$ represents the effective QCD vacuum \cite{Callan}
in the basis where all quark masses are real.
$\overline{\theta}$ can attain any value between 0 and $2\pi$
and a nonzero $\overline{\theta}$ value would imply CP violation.
An observable evidence of CP violation would be an electric dipole moment of neutrons, 
which has theoretically been estimated \cite{Baluni} to
$d_{n} \approx 5 \times 10^{-16} \overline{\theta}$ e cm.
The present experimental upper limit \cite{Ramsey}, however, is smaller than 
$d_{n} < 3 \times 10^{-26}$ e cm.
Consequently, the arbitrary parameter $\overline\theta$ should be smaller than $10^{-10}$. 
Are there physical reasons for $\overline\theta$ being fixed so close to zero?
This open question is referred to as the strong CP problem.\\ 

An elegant solution of the strong CP problem has been proposed by Peccei and Quinn \cite{Peccei} 
in 1977 by introducing an additional global chiral symmetry U(1)$_{PQ}$, 
which is known as PQ-symmetry.
This symmetry is spontaneously broken at an unknown scale $f_{a}$. 
A new, light, and neutral pseudoscalar particle could thus arise, namely an axion \cite{Weinberg}
associated with the spontaneous symmetry breaking. 
The axion field gives an additional term $L_{a}$ in the QCD Lagrangian, and Eq. (1) reads
\begin{eqnarray}
L_{QCD} =  L_{pert} + \overline\theta {g^{2} \over 32\pi^{2}}
G_{a}^{\mu\nu} \widetilde{G}_{\mu\nu}^{a} + L_{a}
\end{eqnarray}
\begin{eqnarray}
L_{a} = -{1 \over 2} (\partial_{\mu} a)^{2} + C_{a} {a \over f_{a}} {g^{2} \over 32\pi^{2}}
G_{a}^{\mu\nu} \widetilde{G}_{\mu\nu}^{a} ,
\label{Laxion}
\end{eqnarray}
where $a$ is the axion field, $f_{a}$ is the scale of the spontaneous PQ symmetry breaking 
and $C_{a}$ is a model dependent constant.
The presence of the second term in Eq. (\ref{Laxion}) provides an effective potential 
for the axion field. At the potential minimum $<a> = -(f_{a}/C_{a})\overline\theta$ 
of this Lagrangian, the CP violating term is eliminated,
so that the strong CP problem is solved in the presence of an axion.\\

All important axion properties depend on the PQ symmetry breaking scale.
The axion coupling to ordinary matter is proportional to the axion mass m$_{a}$
and, equivalently, to the inverse of the PQ scale 1/$f_{a}$.
Axions are generated without mass, however, they get an effective mass by their interaction
with gluons. The axion-gluon vertex induces an axion to quark-antiquark transition, therefore
axions and neutral pions are mixing with each other.
The axion mass has been estimated by Bardeen and Tye \cite{Bardeen} using the current algebra method
and is given by
\begin{eqnarray}
m_{a} = {f_{\pi}m_{\pi^{0}} \over f_{a}}{\sqrt{z} \over 1+z}
= 0.6 {\rm eV} {10^{7} {\rm GeV} \over f_{a}} ,
\label{axion-mass}
\end{eqnarray}
where $m_{\pi^{0}}$ = 135 MeV, $f_{\pi}$ = 93 MeV, and $z = m_{u}/m_{d} = 0.56$
are the pion mass, the pion decay constant, 
and the mass ratio of up and down quarks, respectively.\\

Most experimental searches for axions focus on the axion interaction with two photons
which is described by
\begin{eqnarray}
L_{int} = - {1\over4}{\rm g}_{a\gamma}F_{\mu\nu}\widetilde F^{\mu\nu} a
        = {\rm g}_{a\gamma} {\bf E}\cdot {\bf B} a ,
\end{eqnarray}
where $F$ is the electromagnetic field strength tensor, $\widetilde F$ its dual,
{\bf E} and {\bf B} the electric and magnetic field, and $a$ the axion field.
The axion photon coupling constant is
\begin{eqnarray}
{\rm g}_{a\gamma} = {\alpha \over 2\pi f_{a}} \left[{E \over N}-{2(4+z) \over 3(1+z)} \right] ,
\end{eqnarray}
where $E/N$ is a model dependent parameter \cite{Kim}.
This coupling is usually bound by the energy loss arguments of stars \cite{Raffelt1}
and a constraint from globular cluster stars \cite{Raffelt2} is g$_{a\gamma} \leq 10^{-10}$GeV$^{-1}$.
CAST is sensitive up to this range.
Axions decay to two photons with a mean lifetime $\tau_{a\rightarrow2\gamma} \sim
\tau_{\pi^{0}\rightarrow2\gamma} (m_{\pi}/m_{a})^{5} \approx 10^{24}$ s for 1 eV axion mass.
Hence, axions are quite stable; 
they live longer than the present estimated age of the universe of 10$^{17}$ s.
Therefore, axions would be excellent candidates for cold dark matter and 
possibly constitute some of cosmic dark matter \cite{Kolb}.

\section{The CAST experiment}
Axions could be produced at the core of the sun and other stars 
by interacting with thermal photons in the Coulomb field of electric charges, 
namely by Primakoff \cite{Primakoff} conversion.
The expected solar axion flux on earth is 
$\Phi_{a}=3.67\times10^{11}{\rm cm}^{-2}{\rm s}^{-1}
({\rm g}_{a\gamma}/10^{-10} {\rm GeV}^{-1})^{2}$ with an approximate spectrum \cite{Zioutas}
\begin{eqnarray}
{{\rm d}\Phi_{a} \over {\rm dE}_{a}} = 3.821\times10^{10}
\left( {{\rm g}_{a\gamma} \over 10^{-10} {\rm GeV}^{-1}} \right)^{2} 
{({\rm E}_{a}/{\rm keV})^{3} \over e^{{\rm E}_{a}/1.103{\rm keV}} -1}
{\rm cm}^{-2} {\rm s}^{-1} {\rm keV}^{-1}
\end{eqnarray}
and an average energy of 4.2 keV.
In a transverse magnetic field, the axions can be reconverted 
to X-ray photons which have the energy and momentum of the original axions.
The conversion probability \cite{vanBibber} of an axion to a photon 
in a strong magnetic field is 
\begin{eqnarray}
{\rm P}_{a\rightarrow\gamma} = 1.7\times10^{-17} 
\left( {{\rm B} \cdot {\rm L} \over 9.0{\rm T} \cdot 9.26{\rm m}} \right)^{2}
\left( {{\rm g}_{a\gamma} \over 10^{-10} {\rm GeV}^{-1}} \right)^{2} 
{{\rm sin}^{2}(q{\rm L}) \over (q{\rm L})^{2}} ,
\end{eqnarray}
where B and L are the magnetic field strength and magnet length, respectively,
$q={\rm m}^{2}_{a}/2{\rm E}_{a}$ is the momentum difference between the axion and the photon in vacuum.
Full coherence over the length of the magnet requires $q{\rm L}<<1$.
This constrains the range of axion masses detectable in the CAST experiment to be m$_{a} <$ 0.02 eV.
This is valid for the evacuated superconductive magnet with parameter L being 9.26 m.
The expected number of photons reaching the detector is ${\rm N}_{\gamma} = 
\Phi_{a} \cdot {\rm A} \cdot {\rm P}_{a\rightarrow\gamma} \approx$ 7 events per day 
for g$_{a\gamma}$ = 10$^{-10}$ GeV$^{-1}$ and the magnet bore area A = 14.5 cm$^{2}$.
By filling the beam pipe inside the magnet with a buffer gas, the X-ray photon acquires
an effective mass and the coherence can be restored for a narrow mass range.
The experimental sensitivity can thus be extended to higher axion masses of about 1 eV,
proper gas pressure settings assumed.\\

The first experiment based on the axion helioscope principle proposed by Sikivie \cite{Sikivie} 
was performed at BNL. Later on, the Tokyo axion helioscope \cite{Moriyama} 
with L = 2.3 m and B = 3.9 T obtained g$_{a\gamma} <$ 6.0$\times$10$^{-10}$ GeV$^{-1}$ 
at 95\%C.L. for m$_{a} <$ 0.03 eV in vacuum and 
g$_{a\gamma} <$ 6.8 - 10.9$\times$10$^{-10}$ GeV$^{-1}$  at 95\%C.L. 
for m$_{a} <$ 0.3 eV using a buffer gas with different pressures.
Limits from crystal detectors are much less restrictive
\cite{Avignone}$^{-}$\cite{Bernabei}.\\

The CAST experiment searches for solar axions of energies of a few keV.
The axions from the core of the sun could convert back 
into photons in the 9 Tesla Large Hadron Collider (LHC) prototype superconducting 
magnet \cite{Konstantin}, 
where the two beam pipes of length 9.26 m inside the magnet are straight 
with an effective cross sectional area $2\times14.5$cm$^{2}$.
Each aperture of the bores fully covers the potentially axion-emitting solar core
(about 1/10th of the solar radius).
A full cryogenic system \cite{Barth} is used to cool the superconducting magnet down to 9 T.
The magnet can be moved automatically to follow the sun track. 
The data were taken every morning and evening for approximately 1.5 hours by 
three different kinds of detectors installed at both ends of the dipole magnet, 
which are sensitive in the energy range up to 10 keV.
The time where the magnet is not aligned with the sun is dedicated to background measurements.
The CAST tracking system has been accurately calibrated
by geometric survey measurements; the pointing precision is better than 0.01$^{\circ}$.
A plexiglas Time Projection Chamber (TPC) covering both beam bores looks for photons 
coming from the magnet after axion-to-photon conversion during sunset,
while during sunrise, the photons are detected with 
a MICROMesh GAseous Structure (MicroMegas) \cite{Giomataris}
and a position-sensitive Charge Coupled Device (CCD) \cite{Strueder}, 
installed at the other end of the magnet.
The reason for using different detectors based on complementary techniques is
the possibility of cross check and their systematic effects.
An X-ray focusing mirror telescope designed for the German X-ray satellite mission 
ABRIXAS is mounted between the magnet and the CCD.
It consists of 27 gold-coated mirror shells with a focal length of 1.6 m
and the total efficiency is approximately 35\% in the energy range of 1 to 7 keV.
The X-ray mirror telescope focuses the photons from the magnet bore of 14.5 cm$^{2}$
aperture to a spot size of about 6 mm$^{2}$ on the CCD, 
and thus improving the signal to background ratio by more than 2 orders of magnitude.
First results were obtained from the 2003 data taking \cite{Zioutas}.
Here we present new results from 2004.

\section{Detector improvements for 2004 operation}
All CAST detectors were operated almost throughout the 2004 data collection periods
with improved conditions.
The differences between 2003 and 2004 measurements were mainly
due to an upgraded experimental setup and additional shielding for detectors.
Here the detector improvements for the 2004 operation are described in detail.
The technical aspects of the detectors can be found in Refs.
\cite{Kuster} \cite{Andriamonje}.

\paragraph{TPC :}
In 2004 the TPC improvements focused mainly on two points: a differential pumping 
system for the TPC X-ray windows as well as passive and active detector shielding.
The aim of the differential pumping system is to effectively decrease the gas leaks 
toward the magnet and to reduce the damage to the thin windows due to unexpected
pressure changes or breakdown. By the continuous pumping of an intermediate volume,
the leak rate has been reduced, and thus the total time for data taking during 2004 
is five times more than 2003.
A passive shielding for the TPC was installed in 2004. It consists of an innermost
copper cage of thickness 5 mm, surrounded by a layer of 2.5 cm ancient lead, 
a layer of 1 mm cadmium, and finally a wall of 22.5 cm polyethylene. 
A plastic box covering the full 
setting is slightly pressurized by injection of clean nitrogen to reduce the radon 
contamination from the air in the space close to the detector. 
Moreover, an active shield has been installed on top of the TPC to clearly identify
and reject the muons producing background.
The full detector shielding of the TPC has reduced the average background level by a factor 
2.4 as compared to 2003. In 2004 the background counting rate was about $4\times10^{-5}$ 
counts cm$^{-2}$ s$^{-1}$ keV$^{-1}$ in the energy range of the 
solar axion spectrum.

\paragraph{MicroMegas :}
A new version of the MicroMegas detector and new electronics have been installed 
for the 2004 data taking period.
The new MicroMegas was designed to reduce
the cross talk effect of the read out strips and thus,
the quality of the data is improved.
In addition, a faster VME Digitizing Board, which records the time structure of the 
mesh signal, was installed and has improved the data transfer, reducing the detector dead time. 
An automatic controller has been installed in order to control the calibration source directly
from the data acquisition system to avoid manual operations.
The improved MicroMegas detector reduces the background level to about
$5\times10^{-5}$ counts cm$^{-2}$ s$^{-1}$ keV$^{-1}$ in the energy range of 1 to 8.5 keV. 

\paragraph{CCD :}
In 2004 the CCD detector was operated with an extra shield inside the detector chamber.
The detector shield was modified by adding a 17 to 25 mm thick layer of low activity
ancient lead outside of the copper shielding.
The lead shield inside the CCD chamber is encapsulated in sealed boxes made of high purity copper.
The additional shielding has reduced the background level by a factor of 1.5 as compared
to the previous setup. The differential mean background rate is about 
$7.5\times10^{-5}$ counts cm$^{-2}$ s$^{-1}$ keV$^{-1}$
with a flat distribution over the 1 to 7 keV energy range.
Very important for the CCD setup is the monitoring of the position of the solar spot on the CCD.
An X-ray source based on a pyroelectric crystal producing a peak activity of 100 MBq
is mounted on a manipulator at the TPC end of the CAST magnet more than 10 m away from
the telescope focal plane. The image of this source allows one to monitor the stability
of the alignment of the X-ray telescope and correspondingly to monitor the position of 
the solar spot on the CCD.
In addition, using a parallel laser beam which is focused by the X-ray telescope on to the CCD,
the location of the expected solar axion signal on the CCD can be determined.
From the regularly performed laser and X-ray measurements, the center of the spot was determined
and it was shown to be stable within one pixel (150$\mu$m)$^{2}$ diameter of the CCD.

\section{Data analysis and results}
All detectors were operated with improved conditions
for the whole 2004 data collection period. Each detector acquired data during 1.5 h
per day when the magnet was pointing to the sun. Background measurements
where taken during non alignment periods and have about a 10 fold exposure time.
The data analysis procedures are along the guideline given in Ref. \cite{Zioutas}.
The theoretically expected spectrum for axions in the allowed mass range ($<$ 0.02 eV)
has been estimated and detector and software efficiencies have been applied.
For the TPC and the MicroMegas these spectra which are proportional to g$^{4}_{a\gamma}$
are directly used as fit functions to the experimental 
spectra after background subtraction. For the CCD data collected in 2004
the analysis could be restricted to a small 6 mm$^{2}$ spot on the CCD
since the telescope stability was continuously monitored.
Due to the low counting statistics on the spot 
a likelihood function for the expected signal plus background was required 
instead of using a $\chi^{2}$ analysis.
All the data sets analyzed for the different detectors
are consistent with the absence of a signal and therefore
95\% confidence limits on g$_{a\gamma}$ for each of the data sets have been extracted.
The preliminary results can be combined
by multiplying the Bayesian probability distribution and 
integrating them over the physically allowed region, 
i.e. positive signals,
to find the combined preliminary result for the 2004 CAST data:
\ \\

\begin{figure}[h]
\centerline{\includegraphics[width=0.6\textwidth,angle=0]{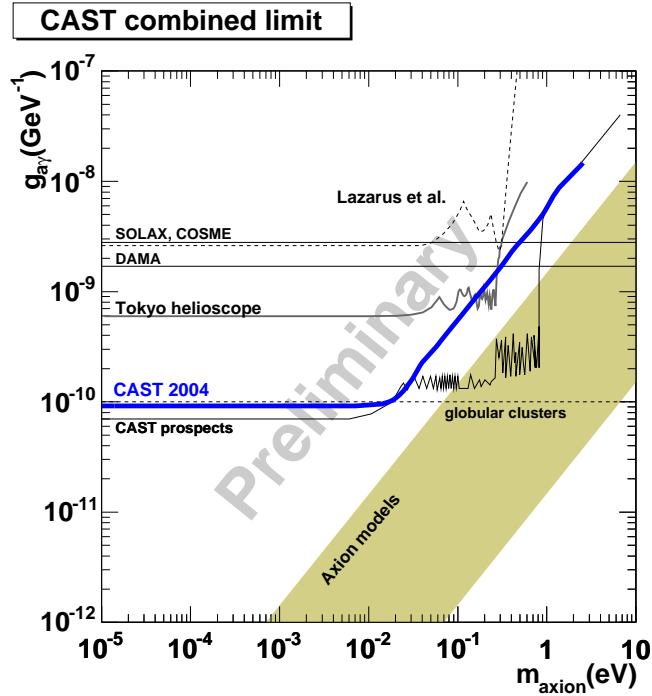}}
\caption{Exclusion limit at 95\% C.L. from the 2004 CAST data compared with constraints 
from other experiments. The shaded region represents the theoretical models.
The future CAST sensitivity is shown as prospects.}
\label{exclusion}
\end{figure}

\begin{center}
g$_{a\gamma} < 0.9 \times 10^{-10}$ GeV$^{-1}$ (95\% C.L.).
\end{center}

\ \\
This result is limited to the axion mass range m$_{a} <$ 0.02 eV where the expected signal 
is independent of the mass since the axion photon oscillation length is larger than
the length of the magnet.
For higher masses, the number of expected signal counts decreases rapidly and 
the shape of the spectral curve differs. 
This analysis procedure was repeated for different values of 
axion masses to derive the whole exclusion line for 95\% C.L. as shown in Figure \ref{exclusion}.   
The systematic uncertainties are estimated to have an effect of less than 
about 10\% on the final results.

\section{Summary and outlook}
The CAST experiment has been operated successfully since May 2003 and during 2004
all three detectors were taking data in upgraded versions with higher sensitivity.
No signal above background was observed in any of the data taken so far.
First results of the 2003 data have been published \cite{Zioutas}.
The preliminary CAST results of the 2004 data yield a lower limit on
g$_{a\gamma} < 0.9 \times 10^{-10}$ GeV$^{-1}$ at 95\% C.L. for axion masses
m$_{a} <$ 0.02 eV. This result improves the previous constraints given by other
experiments by a factor of 7.
It is the first time that an experimental limit falls below the astrophysical limit 
of globular clusters.
So far the CAST experiment has been operated with an evacuated magnet bore
and thus was restricted to axion masses below 0.02 eV.
During 2005 a major modification to the magnet pipe system was undertaken. 
End of 2005 CAST started with measurements for its second phase.
The magnet pipes were filled with a low Z buffer gas (starting with $^{4}$He
and in the future $^{3}$He) of various pressures in order to restore coherence
for higher axion masses m$_{a} >$ 0.02 eV. The extended sensitivity will allow
CAST to reach masses up to 0.8 eV and thus reach into the region of
theoretical axion models.

\end{document}